\documentclass[preprint]{JHEP3}

\usepackage{epsfig,multicol,bbm}

\newcommand\fverb{\setbox\fverbbox=\hbox\bgroup\verb}
\newcommand\fverbdo{\egroup\medskip\noindent%
                        \fbox{\unhbox\fverbbox}\ }
\newcommand\fverbit{\egroup\item[\fbox{\unhbox\fverbbox}]}
\newbox\fverbbox

\title{On the fraction of dark matter in charged massive particles (CHAMPs)}
\author
{F.~J.~S\'anchez-Salcedo\thanks{E-mail:jsanchez@astroscu.unam.mx}, 
E.~Mart\'{\i}nez-G\'{o}mez and J.~Maga\~na\\
Instituto de Astronom\'{\i}a, Universidad 
Nacional Aut\'onoma de M\'exico, \\ Ciudad Universitaria,
Apt.~Postal 70 264, 
C.P. 04510, Mexico City, Mexico}

\received{\today}               
\accepted{\today}               

\preprint{}

\abstract{
From various cosmological, astrophysical and terrestrial requirements,
we derive conservative upper bounds on the present-day fraction of 
the mass of the Galactic dark matter (DM) halo in charged massive 
particles (CHAMPs). 
If dark matter particles are neutral but decay lately into CHAMPs, 
the lack of detection of heavy hydrogen in sea water and the
vertical pressure equilibrium in the Galactic disc turn out
to put the most stringent bounds.  
Adopting very conservative assumptions about the recoiling velocity of
CHAMPs in the decay and on the decay energy deposited in baryonic gas,
we find that the lifetime for decaying neutral DM must be
$\gtrsim (0.9-3.4)\times 10^{3}$ Gyr.  Even assuming 
the gyroradii of CHAMPs in the Galactic
magnetic field are too small for halo CHAMPs to reach Earth,  
the present-day fraction of the mass of
the Galactic halo in CHAMPs should be $\lesssim (0.4-1.4)\times 10^{-2}$.
We show that redistributing the DM through the coupling
between CHAMPs and the ubiquitous magnetic fields cannot be a solution 
to the cuspy halo problem in dwarf galaxies.}


\begin{document}

\section{Introduction}
So far, other than its gravitational interaction, the detailed properties 
of the dark matter (DM) are still largely unknown.
The lack of direct detection suggests that DM particles are
stable, neutral and weakly interacting. However, it is important to know
how well these properties are constrained from an observational point
of view. In recent years, mainly motivated by the cuspy problem of dark
haloes in low-surface brightness galaxies and dwarf galaxies, and the 
excessive abundance of satellite galaxies
inferred in cosmological simulations, much work has been done to study the 
cosmological and astrophysical implications of other variants, 
such as decaying, collisional or annihilating DM.
New interesting phenomena at galactic scales, ignored under the assumptions
of collisionless and neutral DM, arise if a fraction of DM is
made up by massive particles with electric charge (CHAMPs).
Several theoretical physics models beyond the Standard Model (SM) have shown
the possibility of CHAMPs (e.g., \cite{fai07, pos08}).
The existence of CHAMPs is well-motivated
in the model of super Weakly Interacting Massive Particle (super-WIMP)
dark matter \cite{fen03}.
In supersymmetric theories, some popular examples of CHAMPs include a 
slepton, such as the supersymmetric staus (e.g., \cite{fai07, hua06}).

The hypothesis that the DM is made up by a mixture of bare CHAMPs 
and neutraCHAMPs (a neutral bound atom formed by
a CHAMP of electric charge $-1$ and a proton) was considered 
in the late eighties \cite{der90}. 
Dimopoulos et al.\cite{dim90} 
noticed that the astrophysical and terrestrial limits are
hardly compatible with such a scenario
(see \cite{per01} and \cite{tao08} for updated reviews).
Nevertheless, all these constraints were derived for the standard flux
of particles at Earth from the Galactic halo, assuming it was constant
over time.
If DM particles are originally neutral and decay lately into
charged particles, many constraints can be avoided and one must reevaluate 
earlier bounds. 

This paper is organized as follows. In section \ref{sec:assumptions},
we describe our assumptions, which were especially designed to minimize
the potentially disastrous effects of charged DM.
The constraints from big bang nucleosynthesis (BBN) and cosmic microwave
background (CMB) anisotropy on the lifetime of decaying neutral DM particles
are discussed in section \ref{sec:cosmology}. 
In sections \ref{sec:density} and \ref{sec:equilibrium}, 
we examine the distribution of CHAMPs in the disc and
in the halo of the Galaxy. We will discuss the implications of the lack of 
detection of anomalously heavy hydrogen in sea water and
the vertical (magneto-)hydrostatic configuration of the Galactic disc
on the fraction of halo CHAMPs.
Other physical implications of CHAMPs embedded in galaxy clusters are briefly 
discussed in section \ref{sec:clusters}. Concluding remarks are 
given in section \ref{sec:conclusions}.

\section{Assumptions}
\label{sec:assumptions}
In order to remedy purported
problems with the collisionless CDM family of cosmological models on 
galactic and galaxy cluster scales, 
Chuzhoy \& Kolb \cite{chu09} have revived the possibility that a significant
fraction of the DM in haloes is made up by unneutralized CHAMPs. 
These authors claim that the distribution of CHAMPs in galaxies may 
be altered by the coupling between CHAMPs and ubiquitous magnetic fields.  
For instance, CHAMPs may be depleted from the central parts
of galaxies, erasing the DM cusp, if they are accelerated through the Fermi
mechanism in supernova shocks. 

Here we explore a generic scenario in which neutral dark
matter, denoted by $\chi$, decay with lifetime $\tau_{\rm dec}$ into
non-relativistic and exotic massive particles with electric charge. 
The model of Chuzhoy \& Kolb \cite{chu09} corresponds to a scenario where
the decay lifetime is shorter than the age of the Universe. 
The decay of DM into another dark or SM
species could have a bearing upon possible problems with the $\Lambda$CDM
scenario, e.g., the reionization of the Universe,
the structure formation at small-scales and the low abundance of
satellite galaxies, the formation of galactic cores,
the synthesis of light elements in the early Universe,
the origin of ultra-high energy cosmic rays, the positron excess
observed by PAMELA, the Tully-Fisher relation with
$z$ or the gas fraction of galaxy clusters.
However, we will focus on the exclusion range of parameters.

Stringent limits on DM decay into photons or SM particles have been derived
from diffuse $\gamma$-ray observations \cite{kri97}, the
effects on BBN \cite{hol99} and from
the reionization history of the Universe (e.g., \cite{zha07,del09}).
In order to set an upper limit on the fraction of CHAMPs as generous
as possible,
we will consider the most favourable and simplest scenario to permit
the maximum amount of charged particles in galactic haloes.
Our model starts from the following, rather artificial,  assumptions:

(1) The coupling strength of the decay
of a neutral DM particle into two (electrically) charged particles,
$X_{1}^{+}$ and $X_{2}^{-}$, plus a very light (or massless) weak
interacting particle $\bar{X}_{\nu}$, is large.
Therefore, $\chi$ dominantly decays without the emission of photons
or $Z$ bosons. The $\chi$-decay mode is thought to be analogous to
the neutron $\beta$-decay but in the dark sector.
Note that $X_{2}^{-}$ is not the anti-particle of $X_{1}^{+}$ and thus
they may have different masses.
To keep the discussion manageable, however, we will assume 
that both particles have the same mass.

(2) $X_{1}^{+}$ and $X_{2}^{-}$ are stable and cannot decay. This can be
accomplished if they are the lightest particles carrying conserved
'dark baryon' number $B'$ and 'dark lepton' number $L'$, respectively.
Conservation of these quantum numbers in the decay is met
if $\chi$ and $X_{1}^{+}$ are dark baryons and $X_{2}^{-}$ is a dark
lepton and $\bar{X}_{\nu}$ a dark anti-lepton.

(3) The decay $\chi \rightarrow X_{1}^{+}+X_{2}^{-}+\bar{X}_{\nu}$
is the dominant way through
which CHAMPs can be produced.  This non-thermal production of particles
is present in some schemes such as SUSY (e.g., \cite{cer06}).
In practice, it is assumed that DM is neutral before freeze-out
and afterwards it decays into CHAMPs.

(4) The charged $X$ particles from the decay are non-relativistic so
that they are cold DM by the time of structure formation and
its contribution to the reionization of the intergalactic medium
does not contradict CMB data.
This is fulfilled in models where $\Delta m/m_{\chi}\lesssim 2\times
10^{-4}$, where $\Delta m$ is the mass difference between the initial
and final states. Although this fine-tuning is unlikely, it is not
so rare in nature. For instance, the corresponding ratio 
in the classical neutron $\beta$-decay
is comparable ($\sim 8\times 10^{-4}$).

We will see that, even under these somewhat artificial assumptions, 
the fraction of dark matter that can be made by CHAMPS is vanishingly 
small for astrophysical phenomena to be affected and
the standard Cold DM cosmology is recovered.
The proposed scenario is in some ways reminiscent of models
already discussed in the literature but introduce important conceptual
differences. In previous works, decaying DM was introduced 
to dissolve
the central cusp in DM haloes and the overabundance of satellite galaxies
by the depletion \cite{cen01,fer09} or energy release in the 
decays \cite{san03,abd08}.
As suggested by Chuzhoy \& Kolb \cite{chu09}, in the CHAMP model, 
even if the recoiling velocities of CHAMPs were very small,
they may be ejected from the central parts of the galaxies by
Fermi acceleration in shock waves, or from the Galactic disc,
making them very evasive for direct terrestrial detection
(see \cite{chu09} for a discussion).
In the next section, we discuss the limits on $\tau_{\rm dec}$
imposed by BBN and CMB.

\section{Pregalactic constraints}
\label{sec:cosmology}

The standard BBN theory has been well
established to predict precisely the primordial light element
abundances and constrain the number density of long-lived CHAMPs
at $t<10^{5}$ s. 
Since the recombination of CHAMPs with protons, $\alpha$-particles,
electrons, or other CHAMPs to form neutral atoms, occurs well after BBN, 
CHAMPs remain bare at BBN.  
If $X_{2}^{-}$ particles, with masses below the weak scale,
are present at the BBN and they do not decay into other particles, 
excessive production of $^{6}$Li and $^{7}$Li may occur only if
the fractional contribution of negative CHAMPs to the present critical density,
$\Omega_{X}$, is larger than a certain maximum value $\Omega_{X}^{\rm max}
=3\times 10^{-6}$ (e.g., \cite{ham07, jed08}, and 
references therein)\footnote{The 
CHAMP-to-entropy ratio $Y_{X}$ was converted to $\Omega_{X}$ using
$\Omega_{X}h^{2}=2.73\times 10^{11}Y_{X}(m_{X}/1\,{\rm TeV})$.}. 
For masses above the weak scale, this bound can be weakened.
In our case, the abundance of CHAMPs increases
in time due to $\chi$ decays and this constraint only applies at 
$t<10^{5}$ s.
By imposing that, at most, a particle number fraction of  
$\Omega_{X}^{\rm max}/\Omega_{c}$ of $\chi$'s, 
with $\Omega_{c}$ the present density of
cold DM ($\Omega_{c}=0.23$), has decayed
by the end of BBN, $t\simeq 10^{5}$ s, we obtain $\tau_{\rm dec}>200$ yr.

A much more stringent constraint can be placed by studying the
effect on the CMB anisotropy of a scattering
interaction between charged CHAMPs and the photon-baryon fluid.
Throughout the paper, we will refer
to ``charged CHAMPs'' to indicate both the free CHAMPs and the bound
states of CHAMPs having a net electric charge.
Kohri \& Takahashi \cite{koh09} have shown that most of the negative CHAMPs
are captured by $\alpha$ particles, forming charged CHAMPs.
Hence, all the CHAMPs are expected to be in a charged state
at the epoch of recombination.
Since charged particles have low velocities, they can thermalize very quickly
by their interaction with the baryons \cite{der90}.
The effect on the CMB of coupling baryons with charged CHAMPs is equivalent
to a standard model with a larger value of $\Omega_{b}$.
Given that the WMAP uncertainty in $\Omega_{b}$ is $< 3\%$,
the fraction of DM particles that are allowed to decay
before recombination epoch, $t_{R}$, is
$\lesssim 0.03(\Omega_{b}/\Omega_{c})\simeq 6\times 10^{-3}$,
where $\Omega_{b}/\Omega_{c}$ is the fraction of baryonic energy density
$\Omega_{b}h^{2}$,
relative to that of the (cold) DM, $\Omega_{c}h^{2}$. 
Consequently, CMB power spectra is obtained if 
$\tau_{\rm dec}> t_{R}\times 10^{3}/6 \simeq 7\times 10^{7}$ yr, 
where we have used $t_{R}\simeq 376,000$ yr (see \cite{dub04}
for the case of millicharged CHAMPs).

For invisible decay to weakly interacting particles such as neutrinos
or $X_{\nu}$, Gong \& Chen \cite{gon08} constrain 
the decay lifetime to $\tau_{\rm dec}>0.7\times 10^{3}\xi$ Gyr,
where $\xi$ is the fraction of the rest mass which gets converted
to neutrinos or $X_{\nu}$. Our assumption (4) implies
that $\xi \simeq 2\times 10^{-4}$ and, consequently, we should
constrain ourselves to models with $\tau_{\rm dec}>1.5\times 10^{8}$ yr. 

In contrast to the model of De R\'ujula et al.~\cite{der90}, 
the contribution of the pressureless
$\chi$ component, and not neutraCHAMPs, is dominant in driving galaxy 
formation.
In this sense, our model is closer to the standard collisionless CDM.

\section{The density of CHAMPs in the Galaxy} 
\label{sec:density}
After recombination, when photons and baryons become noninteracting,
baryons and charged CHAMPs fall into the gravitational wells formed by 
pressureless $\chi$-particles and neutraCHAMPs. 
In the halo of a galaxy like ours, the temperature becomes high
enough to ionize hydrogen and superheavy hydrogen $(X_{1}^{+}e^{-})$
--see the Appendix \ref{sec:appendixa}--. 
In contrast, although the equilibrium fraction of $(X_{2}^{-}p)$ in the halo
is small, neutraCHAMPs formed at pregalactic stages (we will refer to them
as 'primordial neutraCHAMPs)
 can survive a long time compared to the Hubble time,
without being dissociated (see the Appendix \ref{sec:appendixa}). 
Nevertheless, the initial density of primordial neutraCHAMPs is expected 
to be small \cite{koh09} and they might be
converted to $(X_{2}^{-}\alpha)$ ions by charge-exchange scattering
and then they behave as charged CHAMPs in the lifetime of the
Galaxy \cite{dim90}.

In order to constrain the abundance of CHAMPs in the Galactic disc
and in the halo we must consider the following processes:
(1) magnetic fields can prevent the 
charged CHAMPs in the halo to penetrate the Galactic disc, 
(2) charged CHAMPs may be ejected from the disc and blown either back to the
halo, or right out of the galaxy if charged CHAMPs are shock accelerated
by supernovae (e.g., \cite{dim90, chu09}), 
(3) Neutral $\chi$ particles have no difficulties to penetrate
the disc and may have a decay when they are crossing the disc, 
replenishing the disc with fresh CHAMPs. 
In the following, we study conditions for which the abundance
of CHAMPs is small in the disc, making its detection unlikely, but large 
in the halo.

\subsection{Shielding the disc with magnetic fields}
\label{sec:shielding}
In principle, a charged CHAMP may 
lose all its kinetic energy by interacting with the interstellar gas
as it intersects the Galactic disc 
\cite{bas90}. However,
the penetration of unneutralized coronal CHAMPs along the Galactic disc 
is impeded by the presence of Galactic magnetic fields.  
Chuzhoy \& Kolb \cite{chu09} claim that charged CHAMPs cannot cross the disc 
if they have masses $m_{X}<10^{8}$ TeV.
A more precise calculation taking into account that the Galactic magnetic
field is not plane-parallel is given below.

It is well-known that when charged particles 
interact with a magnetized body, a boundary
layer that divides two regions with different conditions is created 
\cite{par91}.
Thus, charged particles will penetrate this boundary by some
distance before they are turned around by the ${\vec{v}}\times \vec{B}$ 
force (Figure \ref{fig:boundary}). 

\FIGURE{\epsfig{file=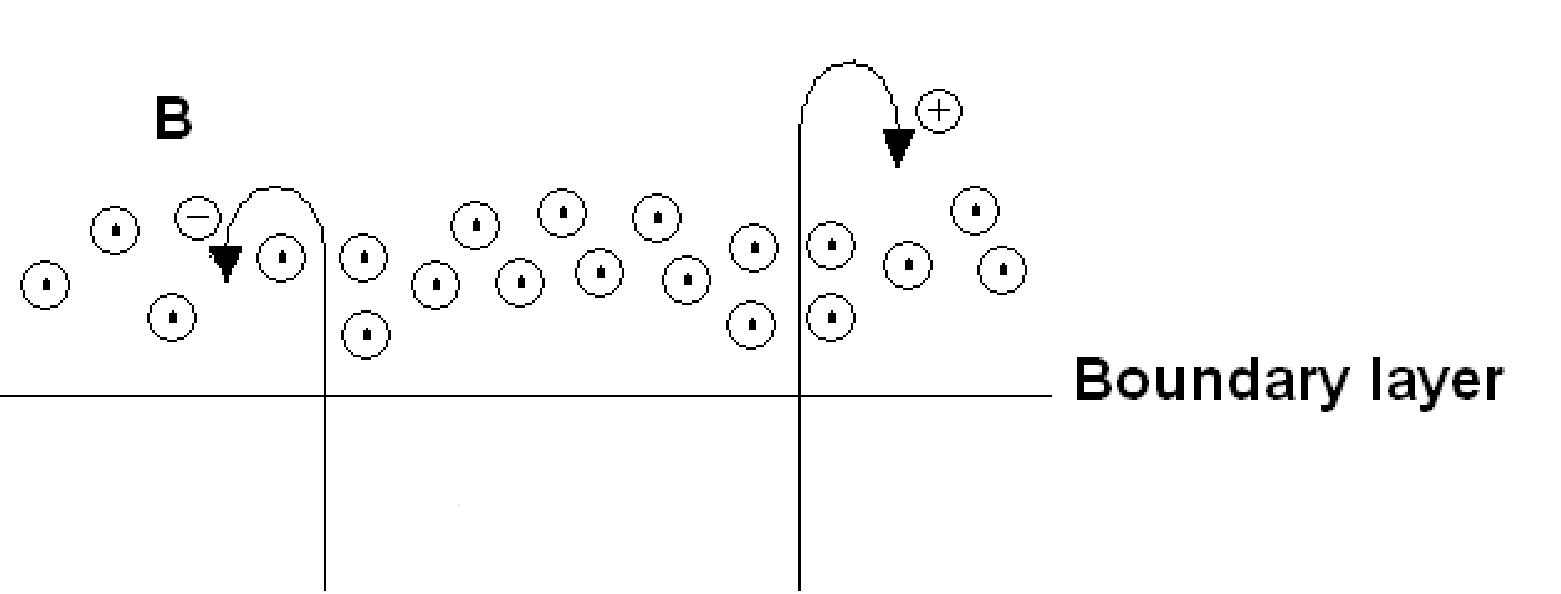,angle=0,width=9.5cm}
  \caption{Sketch of the structure of the magnetic boundary layer 
formed by the partial
penetration of the charged particles before they are deflected back. We also
illustrate the corresponding orbits of particles in the neighborhood of such
boundary for both types of charge, that is, negative and positive.
The degree of shielding depends on the topology of the magnetic field.
In galactic discs, the magnetic field lines are not plane-parallel.
}%
\label{fig:boundary}}

The boundary layer is formed because of the partial penetration
of the charged particles before they are deflected back.
A schematic representation of
the orbits described by negative and positive charged particles in
the neighborhood of the magnetic boundary are drawn in Fig.~\ref{fig:boundary}.

The degree of shielding by magnetic fields depends on the configuration
of the magnetic field.
In the case of the magnetized Galactic disc, the magnetic field is
not plane-parallel and the propagation of
charged particles is more complex because the magnetic field has
a turbulent component, i.e.~$\vec{B}=\vec{B}_{0}+\vec{b}$, 
where $\vec{B}_{0}$ is the regular (homogeneous) magnetic field 
and $\vec{b}$ denotes the turbulent field. The
particle motion is determined not only by the average magnetic field but
also by scattering at field fluctuations, a stochastic process which
requires the solution of transport equations with particle ensembles.
Particle propagation in turbulent fields can be understood as a diffusive
process, reason why we consider the spatial diffusion of 
halo CHAMPs into the galactic disc.

As occurs when one considers the escape
timescale of cosmic rays from the Galactic disc (e.g., \cite{cas02}),
the timescale for the penetration of halo CHAMPs is governed by the 
diffusion timescale across 
the galaxy disc thickness, $\tau_{\rm diff}$. The ordered magnetic lines 
follow the spiral pattern, come out of the Galaxy disc and unfold in the halo. 
However, due to the 
presence of the tangled (turbulent) component of the magnetic field, 
the penetration timescale 
of halo particles through the magnetic spiral arms
is much longer than $\tau_{\rm diff}$ because they must diffuse a distance
much longer than the galaxy disc thickness $H$.
The diffusion timescale $\tau_{\rm diff}$ across $H$
for a halo CHAMP, is bracketted in the range:
\begin{equation}
\frac{H^{2}}{2D_{\parallel}}<\tau_{\rm diff}\lesssim \frac{H^{2}}{2D_{\perp}},
\end{equation}  
where $D_{\parallel}$ and $D_{\perp}$ are the diffusion
coefficients parallel and transverse to the mean component of the magnetic
field, which is observed to be parallel to the disc. 

The magnetic field that feels a charged CHAMP moving within the disc can
be considered static because Alfv\`en waves propagate with velocities of the
order of the Alfv\`en speed $v_{A}\sim 6$ km s$^{-1}$, which is smaller 
than the typical velocities of CHAMPs $\gtrsim \sqrt{3}\sigma_{v}$, where
$\sigma_{v}\simeq 150$ km s$^{-1}$ is the one-dimensional velocity dispersion 
for halo particles.
The diffusion coefficients depend on the turbulence level $\eta\equiv
(1+\left<B_{0}^{2}\right>/\left<b^{2}\right>)^{-1}$, 
and on the rigidity $\chi\equiv 2\pi r_{L}/\lambda_{\rm max}$,
with $r_{L}$ the Larmor radius defined with respect to the total
magnetic field and $\lambda_{\rm max}$ the maximum
scale of the turbulence $\sim H/2$ \cite{cas02, gia99}.
Observations of the Galactic polarized synchrotron background yield 
$1<\left<b^{2}\right>/\left<B_{0}^{2}\right><9$ 
(\cite{fle01}, and references therein),  
implying that $0.5<\eta<0.9$.
Since $\tau_{\rm diff}$ scales as the inverse of the
diffusion coefficients and those are essentially a
monotonic function of $\eta$, we use $\eta\simeq 0.5$ in our
estimate of $D_{\perp}$ in order to give an upper limit on the
diffusion timescale.
From the numerical result by \cite{cas02}, we know 
that $D_{\perp}/(r_{L}v)\sim 0.3$ for Kolmogorov turbulence with
$\eta =0.5$ and $\chi$ between $0.05$ and $0.4$,
we find that the diffusion timescale average over the velocity distribution
is
\begin{eqnarray}
&&\tau_{\rm diff}\lesssim  \left<\frac{5H^{2}}{3r_{L}v_{X}}\right>
=10 \,{\rm Gyr} \left(\frac{H}{300{\rm pc}}\right)^{2}
\left(\frac{m_{X}}{10^{6}{\rm TeV}}\right)^{-1}
\nonumber\\
&&
\times \left(\frac{\sigma_{v}}{150{\rm km \,s}^{-1}}\right)^{-2}
\left(\frac{B}{5\,\mu{\rm G}}\right).
\label{eq:taudiff}
\end{eqnarray}
The values of the turbulence level and rigidity depend on the galactocentric
distance but also on the azimuthal angle in the disc,
because both the regular and the turbulent fields are commonly more intense 
within the spiral arms (\cite{bec07}, and references therein). 
The efficiency of the magnetic shielding can be
reduced along Galactic magnetic chimneys. Moreover,
it is likely that CHAMPs are accelerated to much higher velocities
by supernova shocks as soon as they penetrate inside the disc, decreasing
$\tau_{\rm diff}$ further. 
In the most optimistic scenario where all these effects can be 
ignored, the present configuration and 
strength of the Galactic magnetic field can
prevent diffusion of (unaccelerated) charged CHAMPs across the Galactic disc 
in the lifetime of the disc
for mass particles $m_{X}< 10^{6}$  TeV.
Note that the corresponding gyroradius for a mass of $10^{6}$ TeV
moving at $300$ km s$^{-1}$ in a field of $5\mu$G is $0.2$ pc.
The equation that governs the number of CHAMPs in the disc will 
be discussed in section \ref{sec:seawaters}.

\subsection{Energy gain and loss of CHAMPs in the disc}
\label{sec:gain}
In the foregoing section we have seen that halo CHAMPs
with masses $m_{X}< 10^{6}$ TeV may have difficulty in
penetrating the magnetized Galactic disc, whereas those inside it
would stay confined to the disc unless they are accelerated.  
Charged CHAMPs in the disc
gain energy through electrostatic fields, Fermi acceleration in shock waves, 
and its descendants (e.g., \cite{bla94}),
and lose kinetic energy due to Coulomb scatterings
with electrons and protons of the diffuse interstellar gas.

Consider masses of $m_{X}$ larger than the electron mass\footnote{We will 
not consider the regime $m_{X}<m_{e}$ because they are
excluded for $10^{-15}\lesssim \epsilon <1$, where $\epsilon$ is the electric
charge units of $e_{e}$, the elementary electron charge
\cite{dav00}.}.
The dissipation timescale due to collisions with the electrons
is $\tau_{\rm dis}=E/|\dot{E}|$, with
$E=m_{X}v_{X}^{2}/2$ and 
\begin{equation}
|\dot{E}|=4\pi n_{e}\frac{e_{e}^{4}}{m_{e}v_{X}}\ln\Lambda,
\end{equation}
where $v_{X}$ is the velocity of the CHAMP in the interstellar medium, 
$e_{e}$ is the electron charge, and 
$n_{e}$ is the electron density ($\approx 0.025$ cm$^{-3}$ in the
solar vicinity) and the Coulomb logarithm has a value of
about $20$.  CHAMPs moving in the Galactic disc may avoid strong cooling 
if the dissipation timescale $\tau_{\rm dis}$ 
is greater than the shock acceleration
timescale $\tau_{\rm acc}$, which is $\gtrsim 0.01$ Gyr 
(e.g., \cite{bel78, wan87}). 
The condition $2\tau_{\rm dis}>\tau_{\rm acc}\gtrsim 0.01$ Gyr implies 
that particles with initial velocities
\begin{equation}
v_{X}> v_{\rm crit}\equiv 150\sqrt{3}\,\, {\rm km\,\; s^{-1}} 
\left(\frac{m_{X}}{2\times 10^{3}\,{\rm TeV}}
\right)^{-1/3},
\end{equation}
can be accelerated and escape from the disc, whereas
those particles with velocities $<v_{\rm crit}$ are expected to lose
kinetic energy until they become neutral by recombining with a proton 
to form a neutraCHAMP $(X_{2}^{-}p)$, or an electron to form 
superheavy hydrogen $(X_{1}^{+}e^{-})$. 
Basdevant et al.~\cite{bas90} suggested that neutraCHAMPs and superheavy hydrogen 
in the disc will reach thermal equilibrium with the environment
and will present turbulent
velocities as those of interstellar neutral hydrogen, $\sim 10$ km s$^{-1}$.
Charge-exchange equilibrium with hydrogen dictates that
$20\%-40\%$ of superheavy hydrogen in the disc 
should be ionized. 

Assuming that the velocity distribution of CHAMPs just after the decay
of $\chi$ particles is
Maxwellian, the fraction $F$ of CHAMPs created in the Galactic disc 
that will be trapped in the
disc forming neutraCHAMPs or superheavy hydrogen is
\begin{equation}
F={\rm erf} \left(\frac{v_{\rm crit}}{\sqrt{2}\sigma_{v}}\right)
-\sqrt{\frac{2}{\pi}}\frac{v_{\rm crit}}{\sigma_{v}}
\exp\left(-\frac{v_{\rm crit}^{2}}{2\sigma_{v}^{2}}\right).
\label{eq:ismneutra}
\end{equation}
After one Hubble time, $t_{H}$,
the relative abundance of superheavy hydrogen in the solar neighbourhood is
\begin{equation}
\frac{[X_{1}^{+}e^{-}]}{[{\rm HI}]}=(F\times 10^{-7})
\left(1-\exp\left[-t_{H}/\tau_{\rm dec}\right]\right)
\left(\frac{m_{X}}{2\times 10^{3}{\rm TeV}}\right)^{-1},
\label{eq:ismheavy}
\end{equation} 
where we have assumed a total density of dark matter of $0.01 M_{\odot}$
pc$^{-3}$. 
Searches for interstellar superheavy hydrogen performed
by looking at the Lyman $\beta$ absorption in the direction of
some nearby stars contrain the relative abundance of superheavy hydrogen
over ordinary hydrogen to $<2\times 10^{-8}$ 
(e.g., \cite{bas90, jur82}).
Combining Eqs (\ref{eq:ismneutra}) and (\ref{eq:ismheavy}) using
$\sigma_{v}=150$ km s$^{-1}$, we find
that for $\tau_{\rm dec}\ll t_{H}$, 
the relative abundance
is smaller than $\sim 10^{-8}$ provided that $m_{X}> 6\times 10^{3}$ TeV.
For $\tau_{\rm dec}\simeq t_{H}$, the condition is fulfilled for 
$m_{X}>4.5\times 10^{3}$ TeV.

\subsection{Constraints from sea water searches}
\label{sec:seawaters}

If CHAMPs are singly charged, the most stringent bound on the 
abundance of CHAMPs in the disc
comes from searches of anomalously heavy sea water\footnote{The 
heating of the interstellar H\,{\sc i} gas by the impacts of 
CHAMPs crossing the gaseous disc imposes a constraint of a factor $\sim 5$
less stringent \cite{chi90}.}. 
The gyroradius for a charged CHAMP of mass $2.5\times 10^{3}$ TeV at
$300$ km s$^{-1}$ is $10$ AU in the magnetic field of the solar wind
($50\mu$G in the Earth vicinity). Therefore,  
the arrival of charged CHAMPs at Earth cannot be impeded by the magnetic field
of the solar wind for $m_{X}$ values in the range of interest
($m_{X}\gtrsim 10^{3}$ TeV).
The null results of searches of CHAMPs, between $10^{3}$ TeV and $10^{5}$ TeV, 
in ocean water 
by Verkerk et al.~\cite{ver92} may be used to constrain the admissible
value for $\tau_{\rm dec}$.

Denote by $n^{+}_{h}(R)$ the number density of
positively charged DM particles at galactocentric distance $R$ in the halo. 
Ignoring the expansion of the dark
halo by the recoiling velocities in the decay, $n^{+}_{h}$ increases
in time by:
\begin{equation}
\frac{dn^{+}_{h}}{dt}=-\frac{dn_{\chi}}{dt}=
\frac{n_{\chi}}{\tau_{\rm dec}},
\end{equation} 
where $n_{\chi}(R,t)$ is the number density of neutral DM particles
in the halo. In the equation above, we have neglected the flux of
CHAMPs from the disc to the halo 
since the mass of DM in the disc is small as compared to that
in the quasi-spherical dark halo. 
Solving this equation, we have 
$n^{+}_{h}(R,t)=n_{0}(R)[1-{\rm exp}(-t/\tau_{\rm dec})]$.
Here, $n_{0}(R)$ is the number of $\chi$ particles at the galactocentric
distance $R$ if they would have not decayed. 

As discussed in \S \ref{sec:gain}, for CHAMPs in the disc, 
we need to differenciate between 
initially slow CHAMPs ($v<v_{\rm crit}$) and initially fast 
CHAMPs ($v>v_{\rm crit}$).
Slow CHAMPs lose energy until they reach thermal equilibrium with the 
interstellar medium gas,
whereas fast CHAMPs will be accelerated by supernova shock waves
and eventually be ejected from the disc.
Denote by $n^{+}_{cd}$ the density of 
neutral and ionized superheavy hydrogen 
in the `cold' phase (one-dimensional rms velocities of $\sim 10$ km s$^{-1}$)
and by $n^{+}_{hd}$ the density of  
$X_{1}^{+}$ in the `hot' phase (one-dimensional rms velocities of $\gtrsim 150$ km
s$^{-1}$).
According to our discussion in section \ref{sec:gain}, 
$n_{cd}^{+}$ increases in time according to 
\begin{equation}
n_{cd}^{+}(t)=F n_{0}(R) 
\left[1-\exp\left(-\frac{t}{\tau_{\rm dec}}\right)\right].
\end{equation}
The escape of cold CHAMPs through diffusion is very small and
can be neglected.

For the hot CHAMPs in the disc, we need to consider 
the evacuation from the disc by Fermi
acceleration processes, the replenishment of CHAMPs by the decay
of neutral particles and the flux from the halo
to the disc through diffusion. 
Let $\tau_{\rm esc}$ the timescale for CHAMPs to escape from the disc.
The equation for $n^{+}_{hd}$ at times $t>t_{d}$,
where $t_{d}$ is the epoch of the formation of the magnetized Galactic
disc, is
\begin{equation}
\frac{dn_{hd}^{+}}{dt}= F'\frac{n_{\chi}}{\tau_{\rm dec}}-
\frac{n_{hd}^{+}}{\tau_{\rm esc}}+\frac{n^{+}_{h}-n_{hd}^{+}}
{\tau_{\rm diff}},
\label{eq:nplusd}
\end{equation} 
with $F'\equiv 1-F$.
We have made the approximation that the diffusion of CHAMPs from the halo
to the disc is 
$D_{\perp}\nabla^{2}n^{+}\simeq (n^{+}_{h}-n_{hd}^{+})/\tau_{\rm diff}$. 
Using the value of $n_{\chi}$ previously calculated, Equation
(\ref{eq:nplusd})
can be immediately solved. For instance, if we define $\tau_{\rm eff}$
as 
\begin{equation}
\frac{1}{\tau_{\rm eff}}=\frac{1}{\tau_{\rm esc}}+\frac{1}{\tau_{\rm diff}},
\label{eq:taueff}
\end{equation}
the density of CHAMPs in the disc at $t>t_{d}$ 
is 
\begin{eqnarray}
&&n^{+}_{hd}(t)=n_{0}\frac{\tau_{\rm eff}}{\tau_{\rm diff}}
\left[1+\left(\frac{F'\tau_{\rm diff}-\tau_{\rm dec}}
{\tau_{\rm dec}-
\tau_{\rm eff}}\right){\rm exp}\left(-\frac{t}{\tau_{\rm dec}}\right)\right]
\nonumber\\&&
+C\;{\rm exp}\left(-\frac{t}{\tau_{\rm eff}}\right),
\label{eq:finalnhd}
\end{eqnarray} 
provided that $\tau_{\rm dec}\neq \tau_{\rm eff}$ (see Appendix 
\ref{sec:appendixb} for details). 
The constant $C$ is fixed by imposing the initial condition that
$n^{+}_{hd}(t_{d})=n_{0}F'[1-{\rm exp}(-t_{d}/\tau_{\rm dec})]$.
Since the Galactic disc
is about $12$ Gyr old, we take $t_{d}\simeq 1.6$ Gyr. 

The number density of positive CHAMPs in the sea water is predicted
to be:
\begin{equation}
n^{+}_{{\rm sea}}\simeq \frac{1}{4d}\int_{\rm last\; 3\; Gyr}\!\!\!\!
\!\!\! (n^{+}_{cd}(t)v_{cd}+n^{+}_{hd}(t)v_{hd}) dt,
\label{eq:nplussea}
\end{equation}
where $d$ is the average ocean depth ($d\simeq 2.6$ km), $v_{cd}$ and
$v_{hd}$ are the
characteristic velocities of particles in the frame corotating with the Sun
for the cold and hot population
(i.e.~$v_{cd}\simeq 17$ km s$^{-1}$, $v_{hd}\gtrsim 300$ km s$^{-1}$).
The integration is carried out
over the accumulation time of the CHAMPs inside the sea water, about 
the age of oceans, $\sim 3$ Gyr (e.g., \cite{ver92, kud01}). 

By requiring that the relative abundance of superheavy isotopes of hydrogen,
compared to ordinary hydrogen is less than about $6\times 10^{-15}$
in the range $10$ TeV $<m_{X}<6\times 10^{4}$ TeV
\cite{ver92},
the exclusion diagram was derived for three different values 
of $m_{X}$ (see Figure \ref{fig:diagram}) for $v_{hd}=300$ km s$^{-1}$.
Note that $n_{hd}^{+}\propto v_{hd}^{-1}$. 
For $\tau_{\rm diff}$ we have adopted the maximum value permitted 
by the inequality (\ref{eq:taudiff}). 
Those neutraCHAMPs that are converted to $X_{2}^{-}\alpha$ by charge-exchange
scattering when they cross the disc will also
contribute as a source of heavy hydrogen in Eq.~(\ref{eq:nplussea}).
The inclusion of this additional source of superheavy sea water shifts the
curves in the top panel of Figure \ref{fig:diagram} downwards, making
the allowed region of parameters more restrictive.

The curves in the $\tau_{\rm esc}$ versus $\tau_{\rm dec}$ plane at 
the top panel of Figure \ref{fig:diagram}
define the maximum value of $\tau_{\rm esc}$ compatible with the 
heavy-water searches, as a function of $\tau_{\rm dec}$, i.e.~
$\tau_{\rm esc}^{\rm max}={\mathcal{G}}(\tau_{\rm dec}, m_{X})$.
Along these curves, we have derived the maximum present-day 
values of $n^{+}_{h}$, $n^{+}_{hd}$ and $n^{+}_{cd}$ allowed by
sea water searches (Figure \ref{fig:diagram}b).  
We see that 
in order for CHAMPs to be abundant in the halo (say, $n^{+}_{h}/n_{0}>0.2$),
$m_{X}>10^{4}$ TeV is required.
It is interesting to note that  
the permitted region of parameters $(\tau_{\rm dec},\tau_{\rm esc})$ 
is very restricted for a particle mass of $2\times 10^{3}$ TeV. 

We see from Figure \ref{fig:diagram} that for $m_{X}=2\times 10^{4}$ TeV, 
$\tau_{\rm esc}<0.6$ Gyr
is a guarantee that the sea-water constraint is fulfilled.
However, according to Eq.~(\ref{eq:taudiff}), the population of hot CHAMPs 
should be accelerated to a velocity dispersion of $\gtrsim 3,400$ km s$^{-1}$ 
in order to escape in less than $\sim 1$ Gyr. This value is, of course,
much larger than our reference value $v_{hd}=300$ km s$^{-1}$ used 
to derive $n^{+}_{\rm sea}$ in Eq.~(\ref{eq:nplussea}). 
A set of self-consistent calculations, which include the fact 
that $v_{hd}$ and $\tau_{\rm esc}$
are not independent, shows that solutions compatible with the lack
of detection of sea water and $300<v_{hd}<10,000$ km s$^{-1}$, 
require $\tau_{\rm dec}>500$ Gyr for
a particle mass of $2\times 10^{4}$ TeV and $\tau_{\rm dec}>2.5\times 10^{3}$ 
Gyr for a mass of $2\times 10^{3}$ TeV.

It is now clear that the allowed values for $\tau_{\rm dec}$ depend
on $m_{X}$ but also on the adopted value for $v_{hd}$.
The parameters $\tau_{\rm esc}$ and $\tau_{\rm dec}$ are unconstrained 
for $m_{X}> 6\times 10^{4}$ TeV, because the concentration of 
anomalously heavy hydrogen in sea water is very uncertain.
In the next section we will consider the vertical pressure
equilibrium in the Galactic disc and will find out a more robust
lower limit on the lifetime of $\chi$ particles.

\FIGURE{\epsfig{file=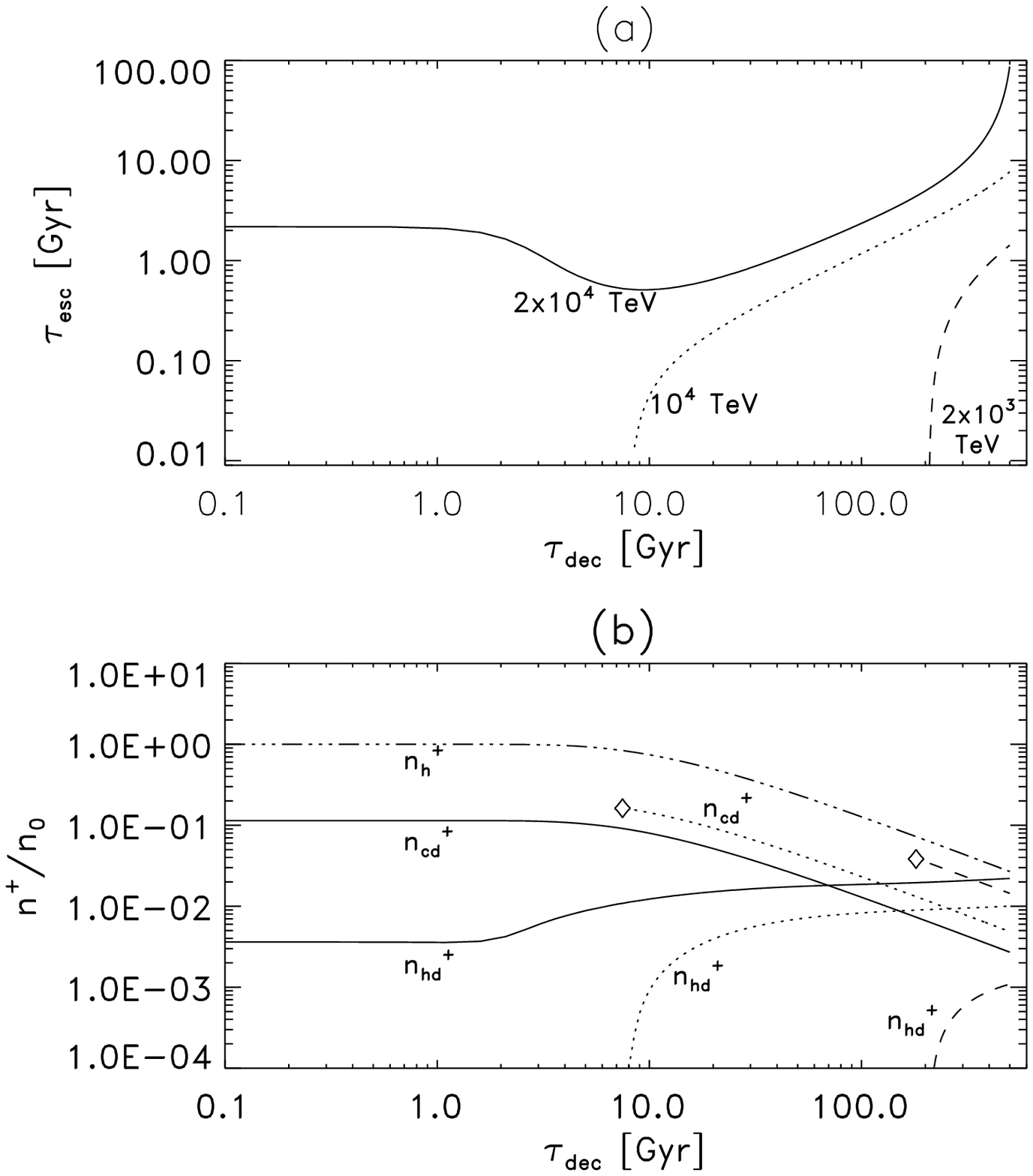,angle=0,width=9.0cm,height=12cm}
  \caption{The allowed regions of $\tau_{\rm esc}$ versus $\tau_{\rm dec}$
parameter space from searches for superheavy hydrogen in deep sea water, 
for different $m_{X}$ (top panel). At each mass $m_{X}$, the region of
parameters below the corresponding curve is allowed. 
The maximum present-day abundance of neutral and ionized superheavy
hydrogen, normalized to $n_{0}$, 
in the halo, cold disc and hot disc, 
allowed by the searches of heavy water in the sea, 
is shown as a function of $\tau_{\rm dec}$ (bottom panel). 
}%
\label{fig:diagram}}

\section{The global magnetic support of halo CHAMPs}
\label{sec:equilibrium}
In the idealised situation that the magnetic field in the disc, at 
$|z|<Z_{\rm min}$, is horizontal
and the halo is unmagnetized,
when charged particles with small gyroradii
try to penetrate the disc, they execute approximately half a 
gyro-orbit before finding themselves back in the unmagnetized region 
and with velocities directed away from the magnetized region
(Fig.~\ref{fig:boundary}).
In the boundary layer, a current
layer develops as a result of a thermal, unmagnetized plasma interacting 
with a magnetized region. 
It is a classical result that the (kinetic) motions of individual 
particles in collisionless 
plasmas can be reconciled with the role inferred for the pressure in MHD
(e.g., \cite{kin67, cra97}). 
Since charged CHAMPs in the halo are essentially collisionless,
the CHAMP momentum flux $m_{X}(n_{h}^{+}+n_{h}^{-})\sigma_{v}^{2}$
at the caps of the disc
should be balanced by the pressure $P_{B}$ of the magnetic field in the disc.
As we discussed in the previous section, the galactic disc
may contain hot CHAMPs that may participate in the pressure balance,
whereas particles trapped in the cold phase do not contribute to support the
halo CHAMPs because its density is very low at $|z|\sim Z_{\rm min}$.
Ignoring the weight of coronal gas and taking $n_{h}^{+}=n_{h}^{-}$
and $n_{hd}^{+}=n_{hd}^{-}$, it holds
\begin{equation}
P_{B}(Z_{\rm min})\gtrsim 2m_{X}n_{h}^{+}\sigma_{v}^{2}-2m_{X}n_{hd}^{+}
\sigma^{2}_{hd}. 
\label{eq:buffi}
\end{equation}
Here $n_{h}^{+}$ is the density at the top of the disc.

Let us now consider the term $n_{hd}^{+}\sigma_{hd}^{2}$ in 
Eq.~(\ref{eq:buffi})
Both CHAMPs and cosmic rays in the Galactic disc may be subject to
Fermi acceleration mechanisms in shock waves. 
It is known that
cosmic rays are in approximate energy equipartition with the magnetic 
field in the diffuse interstellar medium \cite{web98}.
Equipartition arguments are usually adopted to
find the magnetic field in other external galaxies (e.g., \cite{bec96}). 
The concept of equipartition between the magnetic field and 
energetic cosmic rays in our Galaxy (or other galaxies) is consistent 
with the widely held belief that the cosmic rays diffuse through 
the field but do not dominate it. In analogy to cosmic rays, we expect
equipartition between magnetic fields and CHAMPs in the hot disc. 
If we assume that the pressure exerted by the CHAMPs in the hot disc
is a fraction $\beta_{hd}$ of the magnetic pressure, Eq.~(\ref{eq:buffi}) is
simplified to $P_{B}\gtrsim 2(1+\beta_{hd})^{-1}n_{h}^{+}\sigma_{v}^{2}$,
with $\beta_{hd}$ of order of unity. 
It is convenient to express this equation in terms of the mass fraction
of dark matter in charged CHAMPs in the halo $f_{h}$ as
\begin{equation}
P_{B}(Z_{\rm min})\gtrsim \left(\frac{1}{1+\beta_{hd}}\right) f_{h}\rho_{0}\sigma_{v}^{2},
\label{eq:intuitive}
\end{equation}
where $\rho_{0}$ is the density of dark matter 
at the top of the disc if it would have not decayed.
We see that by requiring that the confinement of the magnetic pressure
is entirely due to the halo charged CHAMPs, an upper value on the
abundance of CHAMPs can be derived.

In a real galaxy, the topology of the magnetic field is more complex.
The halo is also magnetized and the magnetic field may rise above the disc.
In the following, we derive a constraint analogous to Eq.~(\ref{eq:intuitive})
but for this more general case.
Suppose that charged halo CHAMPs cannot penetrate
down to $z=Z_{\rm min}$ (with $Z_{\rm min}>H$) in the lifetime of the Galaxy
because of the Galactic magnetic barrier 
(see \S \ref{sec:shielding}). 
Integrating the equation of vertical
equilibrium from $z=Z_{\rm min}$ to $z=\infty$, assuming that the
magnetic field is horizontal at $Z_{\rm min}$ and zero pressure
at $z=\infty$, we find 
\begin{equation}
P_{B}(Z_{\rm min})\gtrsim \frac{1}{1+\beta_{hd}}
\int_{Z_{\rm min}}^{\infty} 
f_{h}\rho_{0}(R,z) K_{z} dz, 
\label{eq:graveq}
\end{equation}
where $\rho_{0}(R,z)$ 
is the density of dark matter if it would have not decayed
and $K_{z}$ the vertical positive gravitational
acceleration. 
Again, Eq.~(\ref{eq:graveq}) is written as an inequality because 
the weight of coronal gas has been neglected.
If the magnetic field is not horizontal at $Z_{\rm min}$ because the
topology of the magnetic field is such that only in certain areas
the magnetic field rises above the disc, the effective vertical
magnetic pressure can be represented by $(B^{2}-2B_{z}^{2})/8\pi$
with $B^{2}$ and $B_{z}^{2}$ interpreted as averages in the $(x,y)$ plane
\cite{bou90}. Therefore, the magnetic tension reduces the effective
vertical magnetic pressure and the inequality in Eq.~(\ref{eq:graveq}) 
still applies.

For a spherical dark halo, the weight term of charged halo CHAMPs  
at the solar vicinity is
\begin{eqnarray}
&&\int_{Z_{\rm min}}^{\infty} f_{h}\rho_{0} K_{z} dz= 
1.7f_{h}\times 10^{-10}{\rm dyn}\,\, {\rm cm}^{-2}
\nonumber \\ &&
\times \left(\frac{\rho_{0}}
{0.01{\rm M}_{\odot}{\rm pc}^{-3}}\right)
\left(\frac{\sigma_{v}}{150\;{\rm km} \;{\rm s}^{-1}}\right)^{2}.
\end{eqnarray}

Interestingly, the observed synchrotron emission
above the plane in the solar neighbourhood implies that the scale height
of the magnetic field is greater than what would be inferred from 
the weight distribution of the interstellar matter (e.g., \cite{cox05}).
The observed synchrotron
emission above the plane in the solar neighbourhood indicates that
the total magnetic field strength is
$2-5\, \mu$G at a height of $z=1$ kpc 
\cite{cox05, fer01, gae08}. 
If we identify $Z_{\rm min}$ as the half width at half maximum
(HWHM) of the magnetoionic disc
$\sim 1$ kpc (e.g., \cite{kal03}) and
by evaluating the magnetic pressure at $z=Z_{\rm min}\approx 1$ kpc,
we obtain the desired constraint on $f_{h}$, once
adopting the highest magnetic value of $5\mu$G allowed by observations:
\begin{equation}
f_{h}\leq  7\times 10^{-3}(1+\beta_{hd}) 
\left(\frac{\rho_{0,\odot}}
{0.01{\rm M}_{\odot}{\rm pc}^{-3}}\right)^{-1}
\left(\frac{\sigma_{v}}{150\;{\rm km}\;{\rm s}^{-1}}\right)^{-2}.
\label{eq:solar}
\end{equation}
This estimate is very robust to 
the precise value adopted for $Z_{\rm min}$ because the magnetic field
decays very slowly with $z$.

In our derivation, we have assumed that the halo is spherical. 
Consider now an oblate isothermal dark halo with axis ratio $q$:
\begin{equation}
\rho_{0}(R,z)=\frac{v_{c}^{2}}{4\pi G\nu q}
\left(R^{2}+\frac{z^{2}}{q^{2}}\right)^{-1},
\end{equation}
where $v_c$ is the asymptotic circular velocity at the equatorial plane
and $\nu=\gamma^{-1}\arcsin \gamma$, with $\gamma=\sqrt{1-q^{2}}$.
In this model, the velocity dispersion is given, within
less than $10\%$, by 
$\sigma_{v}\simeq 1.16 \sqrt{q}(v_{c}/\sqrt{2})$, for flattening $0.05<q<0.5$ 
(e.g.~\cite{ger96}).
Even though the velocity dispersion for $q<1$ is smaller than
in the spherical case, the weight term 
changes only by $\sim 10\%$ as compared to the spherical case,
even for rather flattened haloes ($q\approx 0.5$).

Consider now a portion 
of the disc at larger galactocentric distances, say $R=2R_{\odot}$.
Following the same procedure as in the solar neighbourhood,
we need to estimate the total magnetic pressure at 
$(2R_{\odot},Z_{\rm min})$, 
which should be responsible to give support
to the charged halo CHAMPs.  The large-scale magnetic field
may have a scaleheight $5$--$10$ times the scaleheight of the neutral
gas disc, so that we may assume that $B_{0}(Z_{\rm min})\simeq
B_{0}(z=0)$. 
The random magnetic field is expected to
be roughly in equipartition with the kinetic energy in the turbulence.
Therefore,
its vertical scaleheight should be similar to that of the gas. If
magnetic fields are still a barrier for halo CHAMPs,
then we may assume that $Z_{\rm min}>H$
and, consequently, the magnetic pressure by the random component
at $Z_{\rm min}$ is less than $10\%$ the pressure by the random field
at $z=0$. Collecting both contributions, we derive an
upper limit for the total magnetic pressure
at $Z_{\rm min}$:
\begin{equation}
P_{B}<\frac{B_{0}^{2}+0.1b^{2}}{8\pi}
=\frac{B_{0}^{2}}{8\pi}(1+0.1\alpha),
\end{equation}
where $\alpha\equiv b^{2}/B_{0}^{2}$, with $b^{2}$ and 
$B_{0}^{2}$ evaluated at $z=0$.
The ordered magnetic field is difficult to measure in the outer
Galaxy, but there is evidence that it decays with radius $R$
as a power-law between
$R^{-1}$ and $R^{-2}$, probably as $\exp(-R/R_B)$ with $R_{B}=8.5$ kpc
\cite{hei96, han06}.
The uniform magnetic field in the  solar neighbourhood is
$2$--$4\mu$G, depending on the authors \cite{han06, bec02}.
If we generously take a value in the solar circle of $4\mu$G,
we infer a strength of $B_{0}\sim 1.5\mu$G at $2R_{\odot}$.
Assuming a spherical dark halo with a mass density at $2R_{\odot}$ of 
$\sim \rho_{0,\odot}/4$, then 
$\int f_{h}\rho_{0} K_{z} dz 
=f_{h}\rho_{0,\odot}\sigma_{v}^{2}/4$. 
At $2R_{\odot}$, our assumption that the halo
is spherical is a very good approximation (e.g., \cite{bel06,
fel06}).
By imposing pressure balance at $z=Z_{\rm min}$ (Eq.~\ref{eq:graveq}),
the following constraint for $f_{h}$ is inferred
\begin{eqnarray}
&& f_{h}\leq 2\times 10^{-3} (1+0.1\alpha)(1+\beta_{hd})\nonumber \\
&&
\times \left(\frac{\rho_{0,\odot}/4}
{0.0025{\rm M}_{\odot}{\rm pc}^{-3}}\right)^{-1}
\left(\frac{\sigma_{v}}{150\;{\rm km}\;{\rm s}^{-1}}\right)^{-2}.
\label{eq:twicesolar}
\end{eqnarray}
Other observational estimates assure our
generously-taken magnetic intensity.  In fact,
data from rotation
measurements of pulsars suggest uniform magnetic fields of
$\sim 0.7\,\mu$G at $R=2 R_{\odot}$ \cite{ran94}, which coincides with
the extrapolation of the fit of radial variation of the regular field 
by Han et al.~\cite{han06}.

Beyond $2R_{\odot}$ it is uncertain if supernovae shocks are able
to clean the disc from CHAMPs. It might be also possible
that beyond the optical radius, 
the magnetic field is too weak to prevent CHAMPs from crossing the disc,
but any more complicated analysis is useless in the face of such
ignorance.  

Combining Eqs (\ref{eq:solar}) and (\ref{eq:twicesolar}) and taking
$\beta_{hd}\approx 1$, 
the present-day fraction of charged CHAMPs in the halo must be
smaller than $(4-14)\times 10^{-3}$, which implies 
$\tau_{\rm dec}\gtrsim (0.95-3.4)\times 10^{3}$ Gyr. 
This lower limit on $\tau_{\rm dec}$, which is valid for $m_{X}$
as large as $10^{6}$ TeV, is comparable to the constraint
inferred from the lack of anomalously heavy water in the sea
for $m_{X}\approx 2\times 10^{3}$ TeV (see section \ref{sec:seawaters}). 
We conclude that although charged particles 
can be suspended in the halo, so that they 
would be impossible to detect as they never reach the Earth, 
the mass fraction of charged CHAMPs in the halo
must be rather small.

We must note that, whereas constraints from BBN and heavy-water searches
are only relevant if CHAMPs are singly charged, because for other charges,
a CHAMP no longer behaves as a proton, 
the constraint $f_{h}\lesssim (0.4-1.4)\times 10^{-2}$ derived in this section, 
is independent of charge $\epsilon$, as long as charged
CHAMPs and magnetic fields are in pressure equipartition in the Galactic disc.

\section{Ram pressure stripping and collisions of galaxy clusters} 
\label{sec:clusters}
Magnetic fields couple charged CHAMPs with themselves and with ordinary matter.
This coupling might 
cause ram pressure stripping of both baryonic and DM of 
subhaloes and satellite systems.
Consider, for instance, the collision of two galaxy clusters.
Estimates for the magnetic field strength in clusters range from roughly
$1-10\,\mu$G at the center and $0.1-1\,\mu$G at a radius of $1$ Mpc.
With these values, the ratio between thermal pressure $P_{th}$ and
magnetic pressure for the CHAMPs is 
$\beta\equiv 8\pi P_{th}/B^{2}\approx 2f\times 10^{3-4}$, that is, a hot plasma.
Even in this dynamically weak magnetic field,
the mean gyroradius for a CHAMP with $m_{X}=10^{6}\epsilon$ TeV,
is $\lesssim 5$ pc at the center and 
$\lesssim 50$ pc at $1$ Mpc.
The governing equations of collisionless hot plasmas were developed
by Chew et al.~\cite{che56}, whose theory is known as the Chew-Goldberger-Low
approximation. This approximation, which leads to MHD equations with
anisotropic pressure, is satisfactory when the Larmor frequency is
large compared to other characteristic frequencies of the problem and
the mean particle gyroradius is short compared to the distance
over which all the macroscopic quantities change appreciably
(e.g., \cite{spi62, sch66}).
Therefore, charged massive particles in the halo of galaxy clusters  
can be described in the
fluid-like anisotropic MHD approximation; in the merger process,
they would behave as a clump of fluid, experiencing ram pressure 
stripping and drag deceleration similar to the gas component.
Since CHAMPs should be attached to the gas component,
the observed offset between the centroid of DM
and the collisional gas of the subcluster
in the Bullet Cluster implies $f\ll 1$
(e.g., \cite{nat02, mar04}). 
Although the current lensing data accuracy is not sufficient to derive
the mass distribution of the subcluster in the Bullet Cluster, the derived mass
estimates of the subcluster leave little room for DM in the gas bullet.

Galactic halo CHAMPs 
may also exert ram pressure on the gas component of the LMC and its stream
due to their continuous scattering by
the intrinsic magnetic field of the LMC and the Magellanic stream. 
For a Milky Way-type halo of $\sim 10^{12}$ M$_{\odot}$, 
a fraction $f_{h}$ of $4\times 10^{-3}$
implies that the mass in charged CHAMPs could be up to $\sim 4\times 10^{9}$
M$_{\odot}$ and the density at $50$ kpc of $1.2\times 10^{-6}$ M$_{\odot}$
pc$^{-3}$.  Since these values are smaller
than those required to explain the mass and extension of the Magellanic
Stream and the size and morphology of the gaseous disc of LMC
\cite{mas05},
we cannot reduce any further our upper limit on $f_{h}$
with the current observations of the LMC disc and the Magellanic Stream.

\section{Conclusions}
\label{sec:conclusions}

Whilst the common wisdom holds that DM is neutral and collisionless,
it is important to explore the possibility of it having nonzero,
not necessarily integer, charge.
If a fraction of the mass of haloes is made up by charged CHAMPs, it may have a 
strong impact on the observable Universe because of the coupling
between magnetic fields and CHAMPs. For instance, ejection of
charged CHAMPs from the regions with intense magnetic fields,
i.e., from the central parts of galaxies, would help alleviate the cuspy halo
problem. In this work, we have constrained the {\it present} abundance of 
CHAMPs in galactic haloes. 

We have explored a model where neutral dark matter decays into non-relativistic
charged products. From BBN and CMB, we find that the decay lifetime 
should be $\gtrsim 0.1$ Gyr. The non-detection of heavy sea-water
puts a limit on the timescale for charged CHAMPs to escape from the 
Galactic disc.
We have considered the pressure support of CHAMPs 
in our Galaxy to derive a simple, upper limit
on the fraction of CHAMPs and milliCHAMPs in galactic haloes. 
Assuming that the accelerated CHAMPs in the disc are in pressure 
equilibrium with the magnetic field,
we find that $f_{h}\lesssim (0.4-1.4)\times 10^{-2}$. 
This constraint rules out CHAMPs as the origin of the cores in LSB
and dwarf galaxies.
The reduction of the central density 
after they have driven the formation of galactic haloes would be
insignificant. Even if all the CHAMPs were depleted from
the central parts of the galaxies, the rotation velocity in a certain
galaxy would suffer a negligible change of $(0.2-0.7)\%$ for
$f_{h}\sim (0.4-1.4)\times 10^{-2}$. 
In the range of astrophysical interest,  
CHAMPs behave like strongly interacting (fluid-like) dark matter (SIDM).
Thus, they face many of the problems attributed to SIDM.
As some examples, we have discussed the survival of the Magellanic Stream and 
the mass distribution of the
Bullet Cluster.
Our constraint that the mass in CHAMPs in the Galaxy is not larger than the
mass of coronal gas in the halo seems to apply also to galaxy clusters.

\section*{Acknowledgments}
We thank Ricardo L\'opez, Leonid Chuzhoy, Gustavo Medina-Tanco, and
Julio Martinell for valuable comments which
led to significant improvements in the manuscript.
F.J.S.S.~acknowledges financial support from PAPIIT project
IN114107 and CONACyT 2006-60526.
E.M.G.~thanks support from DGAPA-UNAM postdoctoral fellowship.

\appendix
\section{Recombination and ionization of CHAMPs in the Galactic corona}
\label{sec:appendixa}
Positive CHAMPs can recombine with free electrons in the Galactic
corona to form neutral bound atoms $(X_{1}^{+}e^{-})$. 
As $X_{1}^{+}$'s behave exactly
like protons with thermal velocity dispersion\footnote{CHAMPs and protons 
have different temperatures. The thermal (one-dimensional) velocity dispersion
of protons of the hot gas ($\sim 10^{6}$ K) at the Galactic corona is
$v_{p} =\sqrt{k_{B}T/m_{p}}\approx 100$ km s$^{-1}$, which
is a bit smaller than the adopted value for the velocity dispersion
of dark matter particles in the halo ($\sim 150$ km s$^{-1}$).}
of $150$ km s$^{-1}$,
the fraction of $X_{1}^{+}$ nuclei that are neutral is comparable, or even smaller,
than the neutral fraction of hydrogen in the coronal gas,
which is extremelly small ($\sim 10^{-6}$, e.g., \cite{spi78}),
and therefore it can be ignored.

The recombination rate of negative CHAMPs with protons in the Galactic
halo including the ground $n=1$ level is given by
$\alpha_{R}^{(1)} n_{p}n^{-}_{h}$, where the coefficient for
recombination is:
\begin{equation}
\alpha_{R}^{(1)}=\left<\sigma_{\rm rec} v\right>
=\frac{2^{10}\exp(-4)\pi \sqrt{\pi}\alpha^{3}h^{2}}{3m_{p}^{2}v_{\rm eff}}
\sum_{n=1} \frac{1}{4n^{2}},
\end{equation}
where $v^{2}_{\rm eff}=v_{X}^{2}+v_{p}^{2}$ (e.g., \cite{der90, fen09}).
At the Galactic halo, we have $v_{\rm eff}\simeq 180$ km s$^{-1}$ and thus
$\alpha_{R}^{(1)}=4\times 10^{-18}$ cm$^{3}$ s$^{-1}$.

Ionization of $(X_{2}^{-}p)$ by collisions with $X_{2}^{-}$ and $X_{1}^{+}$
are also important for the determination of the abundance of
neutraCHAMPs in galactic haloes.
The coefficient for collisional ionization is:
\begin{equation}
\gamma_{\rm ion}=1.3\times 10^{-8} T_{X}^{1/2} F E_{\rm bin}^{-2}({\rm eV})
\exp\left(-\frac{E_{\rm bin}}{k_{B}T_{X}}\right) [{\rm cm}^{3}/{\rm s}],
\end{equation}
where $E_{\rm bin}\simeq 25$ keV is the binding energy of the atom 
$(X_{2}^{-}p)$,
$T_{X}$ the temperature of bare CHAMPs and $F\simeq 0.83$ for hydrogenic
atoms (e.g., \cite{cox69}).
For massive CHAMPs in the Galactic halo, we have
\begin{equation}
\gamma_{\rm ion}=
1.0\times 10^{-10} \left(\frac{m_{X}}{2\times 10^{4}{\rm TeV}}\right)^{1/2}
\left(\frac{\sigma_{v}}{150 {\rm km/s}}\right) [{\rm cm}^{3}/{\rm s}].
\end{equation}
Putting together,
\begin{equation}
\frac{dn^{-}_{h}}{dt}=\frac{n_{\chi}}{\tau_{\rm dec}}+
\gamma_{\rm ion}(n^{-}_{h}+n^{+}_{h})n_{nC}
-\alpha_{R}^{(1)} n_{p}n^{-}_{h},
\end{equation}
where $n_{nC}=n^{+}_{h}-n^{-}_{h}$ is the density of neutraCHAMPs
in the halo.
The abundance of neutraCHAMPs as compared to CHAMPs at ionization
equilibrium is 
$\sim 1.0\times 10^{-3}$. This estimate should be considered as
an upper limit because $(X_{2}^{-}p)$ may be converted to
a charged state, $(X_{2}^{-}\alpha)$, by a charge exchange reaction.
Unfortunalely, the exchange cross section is very uncertain \cite{dim90}.

Coronal neutraCHAMPs can penetrate the Galactic disc, reach Earth and stop 
in the atmosphere or ocean \cite{dim90}.
Searches for coronal neutraCHAMPs in cosmic rays rule out particles with masses
between $100$ and a few $10^{4}$ TeV if all the $X_{2}^{-}$ are bound to
a proton and the charge exchange cross section with C and O nuclei
is in the interval from $30$ mb to $30$ b \cite{bas90, bar90, hem90}.

\section{Solving the differential equation for $n_{hd}^{+}$}
\label{sec:appendixb}
In section \ref{sec:seawaters}, we derived the differential equation for the 
number density of hot $X^{+}$ in the disc as:
\begin{equation}
\frac{dn_{hd}^{+}}{dt}= F'\frac{n_{\chi}}{\tau_{\rm dec}}-
\frac{n_{hd}^{+}}{\tau_{\rm esc}}+\frac{n^{+}_{h}-n_{hd}^{+}}
{\tau_{\rm diff}},
\label{eq:init}
\end{equation} 
with
\begin{equation}
n_{\chi}(t)=n_{0}\exp\left(-\frac{t}{\tau_{\rm dec}}\right),
\end{equation}
and
\begin{equation}
n_{h}^{+}(t)=n_{0}\left[1-\exp\left(-\frac{t}{\tau_{\rm dec}}\right)\right].
\end{equation}
Inserting them into Eq.~(\ref{eq:init}), we find
\begin{equation}
\frac{dn_{hd}^{+}}{dt}+\frac{n_{hd}^{+}}{\tau_{\rm eff}}
=\frac{n_{0}}{\tau_{\rm diff}} g(t),
\label{eq:firstorder}
\end{equation} 
where the definition of $\tau_{\rm eff}$ was given in Eq.~(\ref{eq:taueff})
and 
\begin{equation}
g(t)\equiv 1+\left(F'\frac{\tau_{\rm diff}}{\tau_{\rm dec}}-1\right)
\exp\left(-\frac{t}{\tau_{\rm dec}}\right).
\end{equation}
The general solution of Eq.~(\ref{eq:firstorder}) when 
$\tau_{\rm dec}\neq \tau_{\rm eff}$ is
\begin{equation}
n_{hd}^{+}(t)=\exp\left(-\frac{t}{\tau_{\rm eff}}\right)
\left[\int \frac{n_{0}}{\tau_{\rm diff}} g(t) 
\exp\left(\frac{t}{\tau_{\rm eff}}\right) dt+C\right],
\end{equation}
where $C$ is a constant. The integral can be performed analytically:
\begin{equation}
\exp\left(-\frac{t}{\tau_{\rm eff}}\right)\int g(t) \exp\left(\frac{t}{\tau_{\rm eff}}\right) dt=
\tau_{\rm eff} \left[1
+\left(\frac{F'\tau_{\rm diff}-\tau_{\rm dec}}
{\tau_{\rm dec}- \tau_{\rm eff}}\right)
\exp \left(-\frac{t}{\tau_{\rm dec}}\right)\right].
\end{equation}
The resulting form for $n_{hd}^{+}(t)$ is given 
in Eq.~(\ref{eq:finalnhd}).

\end{document}